# Electromechanical Piezoresistive Sensing in Suspended Graphene Membranes

A.D. Smith[1], F. Niklaus[1], A. Paussa[2], S. Vaziri[1], A.C. Fischer[1], M. Sterner[1], F. Forsberg[1], A. Delin[1], D. Esseni[2], P. Palestri[2], M. Östling[1], M.C. Lemme[1,3,*]

[1] KTH Royal Institute of Technology, Isafjordsgatan 22, 16440 Kista, Sweden,

[2] DIEGM, University of Udine, Via delle Scienze 206, 33100 Udine, Italy

[3] University of Siegen, Hölderlinstr. 3, 57076 Siegen, Germany

[*] Corresponding author: max.lemme@uni-siegen.de

**Abstract**

Monolayer graphene exhibits exceptional electronic and mechanical properties, making it a very promising material for nanoelectromechanical (NEMS) devices. Here, we conclusively demonstrate the piezoresistive effect in graphene in a nano-electromechanical membrane configuration that provides direct electrical readout of pressure to strain transduction. This makes it highly relevant for an important class of nano-electromechanical system (NEMS) transducers. This demonstration is consistent with our simulations and previously reported gauge factors and simulation values. The membrane in our experiment acts as a strain gauge independent of crystallographic orientation and allows for aggressive size scalability. When compared with conventional pressure sensors, the sensors have orders of magnitude higher sensitivity per unit area.

Keywords: graphene, pressure sensor, piezoresistive effect, nanoelectromechanical systems (NEMS), MEMS

Graphene is an interesting material for nanoelectromechanical systems (NEMS) due to its extraordinary thinness (one atom thick), high carrier mobility [1,2], a high Young's modulus of about 1 TPa for both pristine (exfoliated) and chemical vapor deposited (CVD) graphene.[3,4] Graphene is further stretchable up to approximately 20%.[5] In addition, it shows strong adhesion to $SiO_2$ substrates[6] and is nearly impermeable for gases, including helium.[7] In this article, we demonstrate piezoresistive pressure sensors based on suspended graphene membranes with direct electrical signal read-out. We utilize a piezoresistive effect induced by mechanical strain in the graphene, which changes the electronic band structure[8] and exploits the fact that the sensitivity of membrane-based electromechanical transducers strongly correlates with membrane thickness[9]. While graphene has been used as a piezoresistive strain gauge on silicon nitride[10] and polymer membranes[11], we extend the use of the graphene to both membrane and electromechanical transduction simultaneously with an average gauge factor of 2.92. The sensitivity per unit area of our graphene sensor is about 20 to 100s of times higher than that of conventional piezoresistive pressure sensors. The piezoresistive effect is nearly independent of crystallographic orientation.

In our experiments, graphene membranes made from CVD graphene are suspended over cavities etched into a $SiO_2$ film on a silicon substrate. The graphene is electrically contacted and the devices are wire-bonded into a chip package. Process schematics are shown in Fig. 1a through c, while details of the fabrication process are described in the methods section. A scanning electron microscope image of a wire-bonded device and a photograph of a packaged device are shown in Fig. 1d. If a pressure difference is present between the inside and the outside of the cavity (compare Fig. 1c), the graphene membrane that is sealing the cavity is deflected and thus strained. This leads to a change of device resistivity due to the piezoresistive effect in the graphene. Measurements were performed in an argon environment in order to reduce the effects of adsorbates. If air is used instead of argon for the experiments, adsorption of non-inert gases and/or molecules on the graphene will affect the resistivity (see details in supporting information).

In the experiments, the packaged devices are placed inside a vacuum chamber. The chamber is then evacuated from atmospheric pressure down to 200 mbar, and then vented back to 1000 mbar. Thus the air sealed inside the cavity presses against the graphene membrane with a force proportional to the chamber pressure. The resistance of the graphene sensor is measured in a Wheatstone bridge (see supporting information), where the graphene membrane is one of the resistors in the bridge. The Wheatstone bridge is balanced at atmospheric pressure by adjusting a potentiometer to the same resistance value as the graphene membrane. The bridge is biased with 200mV square wave pulses with durations of 500 µs. These values were chosen to avoid excessive heating of the graphene device. The voltage output signal from the Wheatstone bridge is amplified and low pass filtered before being sampled with an analog-to-digital converter and converted into its corresponding resistance value. The experimental conditions were chosen to remain within the expected tearing limits of the graphene membrane.[6]

The suspended membrane sensors were first compared to devices with identical dimensions fabricated in parallel, but without cavities. This was done in order to verify that it is indeed the presence of the cavity and the resulting mechanical bending and straining of the membrane that causes the pressure dependence of the resistance.

Fig. 2a shows the amplified voltage (with an amplification factor of 870) versus pressure curve for two devices, one with a cavity and one without. In contrast to the reference device without the cavity (red hollow circles), the sensor device with the suspended graphene membrane (blue squares) shows a strong correlation of the resistance with respect to pressure. The graph includes data of six measurement cycles, where each cycle represents one pump-down or one venting of the chamber. Fig. 2b shows the average change in voltage of three cavity devices in comparison to two non-cavity devices. As can be seen, there is a very strong correlation between the devices' sensitivity to pressure and the presence of a cavity. Finally, A device was held at constant pressures in order to investigate potential drift in the sensor signal (Fig. 2c). While there is a noticeable drift at several pressures, the resistance values generally follow the pressure. Nevertheless, further studies regarding stability are required.

The sensitivity of piezoresitive membrane-based pressure sensors is given by Eq. 1, where $S$ is the sensitivity, $R$ is the resistance, $V$ is the voltage, $I$ is the current, and $P$ is the pressure difference acting on the membrane.[12]

$$S = \frac{\Delta R}{R \cdot P} = \frac{\frac{\Delta V}{I}}{\frac{V}{I} \cdot P} = \frac{\Delta V}{V \cdot P} \qquad (1)$$

If the current is held constant, then the sensitivity based on voltage measurements can be directly compared to the sensitivity based on the maximum change in resistance for a change in pressure of 477 mbar. The sensitivity of the piezoresitive graphene pressure sensor in Fig. 2 is measured to be 3.95 µV/V/mmHg.

The graphene membrane-based pressure sensor, though much smaller than conventional piezoresitive pressure sensors, outperforms conventional piezoresitive Si-based and carbon nanotube (CNT) based pressure sensors reported in literature.[13-16] (see Table 2 in the supporting information for details).

In general, the sensitivity $S$ of membrane-based piezoresistve pressure sensors is dependent on the membrane material characteristics, the membrane thickness and the membrane area (see supporting information).[3,12,17,18] When normalizing the sensitivity of the pressure sensors from Table 2 of the supporting information to a standard membrane area, the sensitivity of our graphene sensor is about 20 to 100s of times higher than the other sensors (Fig. 2d). This and the fact that the graphene sensor is already smaller in area than any of the other sensors indicate great potential for further size-reduction of graphene membrane-based sensors.

In order to estimate the piezoresistive gauge factor of the graphene transducer in our sensor, the change in resistance of the cavity region must be determined. A finite element analysis of the deflection was performed using COMSOL multiphysics and calibrated using literature data of graphene membrane deflection obtained by atomic force microscopy.[6,7] Material parameters taken from literature were used in the COMSOL model such as the elastic constant $Et$ = 347 N/m, where $E$ is the Young's modulus, $t$ is the membrane thickness ($t$ = 0.335 nm) and Poisson's ratio $v$ = 0.16. A comparison between the model and measured literature values is shown in Table 2 of the supporting information. Good agreement is noted both with the measurements of a 2.3 µm radius circular membrane in Koenig et al.[6] and with the measurement in Bunch et al. on a square 4.75 µm x 4.75 µm membrane at a pressure difference of 930 mbar.[7]

The derived model was then applied to the 6 μm by 64 μm cavity used in the current experiment to estimate the deflection of the membrane. At a pressure difference of 477 mbar, the deflection of the membrane is calculated to be 202 nm (Fig. 3a), which results in an average strain of 0.290% across the membrane.

An electrically equivalent circuit of the sensor is schematically shown in Fig. 3b and is described in Eq. 2

$$R_{tot} = R_1 + \frac{1}{\frac{1}{R_2}+\frac{1}{R_4}+\frac{1}{R_5}} + R_3 \qquad (2)$$

The total resistance $R_{tot}$ is taken from resistance measurements at chamber pressures of 1000 mbar and 523 mbar (Fig. 2a). The resistances $R_1$ through $R_5$ correspond to the resistances of the regions shown in Figure 3b. $R_2$ represents the resistance of the graphene membrane over the cavity and we assume that only $R_2$ changes as a function of pressure. Using this method (details in supporting information), $R_2$ is determined to be 0.191 kΩ at 1000 mbar, and the percent resistance change of the graphene membrane patch ($R_2$) is determined to be 0.59%.

The intrinsic graphene gauge factor in our sensor was then calculated as the percent change in resistance divided by the percent change in strain to be 3.67. Gauge factors vary depending on the pressure range measured with a maximum value of 4.33 and an average value of 2.92. Previous literature, by comparison, reports gauge factors of 1.9 for suspended graphene beams[19], about 150 for graphene on $SiO_2$[20] and nearly 18000 for graphene on a silicon nitride membrane.[10]

Simulations of the change of the graphene pressure sensor resistance due to strain were carried out in order to interpret the experimental results. For a low electric field in the transport direction and considering a Fermi level close to the Dirac energy ($E_F = 0$ eV), the resistance R$_2$ of the graphene foil suspended over the cavity is expressed as

$$R_2 = \rho \frac{L'}{W_t'} = \frac{1}{2qN_e(\varepsilon)\mu_e(\varepsilon)} \frac{(1+\varepsilon_{xx})L}{(1+\varepsilon_{yy})W_t}, \quad (3)$$

where $\rho$ is the resistivity of the suspended graphene sheet, $N_e$ and $\mu_e$ are respectively the electron density and the corresponding mobility (that are assumed to be the same as the hole density and hole mobility, respectively, since we set E$_F$ = 0 eV), $\varepsilon_{xx}$ and $\varepsilon_{yy}$ are the components of the strain tensor respectively in the direction of the transport and normal to the transport, and $q$ is the positive electron charge. Note that the strain induced in graphene by the pressure difference between the cavity and the chamber $\Delta p$ influences both the terms $L'$ and $W_t'$ related to the geometry as well as the resistivity $\rho$. The induced strain can be considered quasi-uniaxial since its component in the direction of the transport is dominant ($\varepsilon_{xx} \gg \varepsilon_{yy}$). If $\rho$ is not modified by strain, the change of $L$ and $W_t$ alone is not enough to explain the experimental resistance change with strain. Then, the influence of strain on $\rho$ has been analyzed by starting with the effect of strain on the electron density $N_e$, simulated by employing the strained graphene bandstructure stemming from the Tight-Binding (TB) Hamiltonian presented in Pereira et al.[8], recalibrated to accurately reproduce DFT results reported by Huang et al.[19] (see supporting information). $N_e$ increases with the strain, which, in contrast with the experiments (see Fig. 2a), would lead to a decrease of the resistivity (see Eq. 3 and supporting information). Hence the changes in the graphene charge are not sufficient to explain the observed change of the resistance $R_2$ with the strain. The effect of capacitive coupling was also explored through simulation and, though present, is found to cause changes in resistance much lower than those observed experimentally.

For this reason the effect of the strain on the mobility $\mu_e$ is simulated by solving the Linearized Boltzmann Transport Equation (LBTE)[21]. This approach gives the exact solution of the LBTE even in the presence of anisotropic and non-monotonic energy dispersion relation and anisotropic scattering rates. In these calculations Neutral Defects (ND)[22] are considered, which are dominant in CVD graphene[23]. The electron mobility decreases with increasing $\Delta p$ (see supporting information). Such mobility degradation more than compensates the $N_e$ enhancement, so that the calculations lead to an overall $R_2$ increase. The simulated versus measured $R_2$ modulation versus strain is also compared (Fig. 3c). As can be seen the simulation results do not critically depend on whether the strain is uniaxial or biaxial and the overall agreement with experiments is reasonably good. Fig. 3d compares the corresponding calculated and measured gauge factors; the biaxial or uniaxial nature of the strain has a modest influence on the gauge factor. Simulation data from Huang et al. are also included for reference.[19] Note also that, due to the flexibility of the approach, the relative variation of $R_2$ is independent of the graphene orientation with respect to the direction of the transport (Fig. 3e and 3f), of the defect concentration, and of the considered Fermi level (i.e. the carrier density). This is an important aspect of our work because it means that the effect is independent of random crystallographic alignment and multiple grain graphene flakes.

The piezoresistive effect in graphene was demonstrated in graphene–membrane pressure sensors.  The sensitivity of piezoresistive graphene sensors is superior to silicon and CNT-based sensors and orders of magnitude more sensitive when normalized for membrane dimensions. This is in line with theoretical considerations that indicate such a decisive advantage due to graphene's extraordinary thinness.  A finite element simulation is derived to describe the deflection of graphene membranes over sealed cavities as a function of pressure and verified with literature data. The estimated maximum gauge factor for graphene based on this model is 4.33 with values averaging at 2.92. Tight binding calculations support the experimental data, including only a small dependence of the observed effect on crystal orientation. This work demonstrates that thin graphene membranes can be efficiently implemented as piezoresistive transducer elements for emerging NEMS sensors.

**Methods**

Devices are fabricated on p-type silicon substrates with a thermally grown silicon dioxide ($SiO_2$) layer of 1.5 µm. Rectangular cavities of 6 µm by 64 µm are etched 650 nm deep into the $SiO_2$ using a resist mask and an Ar and $CHF_3$-based reactive ion etching (RIE) process at 200mW and 40 mTorr to provide vertical etch profiles. Next, contact areas are defined by lithography and etched 640 nm into the $SiO_2$ layer using again an RIE process. The contact cavities are then filled with a 160 nm layer of titanium followed by a 500 nm layer of gold using metal evaporation so that the contacts are raised about 20 nm above the surface of the $SiO_2$. The contacts are buried to prevent wire bonding from ripping the contacts off of the substrate. This has the added advantage of allowing the graphene to be transferred in a later step, which improves the cleanliness of the process and reduces the risk of rupturing the graphene membranes during processing. Also, the graphene-metal contacts are not degraded by polymer residues in this way. Commercially available chemical vapor deposited (CVD) monolayer graphene films on copper foils are used. The graphene on one side of the copper is spin-coated with either a poly(methyl methacrylate) (PMMA) or poly(Bisphenol A) carbonate (PC) layer in order to act as a mediator between the initial and final substrate [24-28]. The graphene on the backside of the foil is etched using $O_2$ plasma and the copper foil is subsequently wet etched in $FeCl_3$ and then transferred into de-ionized water. The bottom left of Fig. 2a shows a contrast enhanced image of graphene with a polymer coating floating in a solution of $FeCl_3$ after the copper is etched away. The PMMA/graphene film is picked up with the chip and dried on a hotplate. After drying, the chip is placed into a solution of Chloroform overnight in order to etch the PC polymer layer. Next, a photoresist layer is applied and exposed in order to pattern the graphene. Finally, the graphene is etched into the desired shape using an $O_2$ plasma etch and the photoresist is removed in acetone. Once the devices are fabricated, the chips are placed into a chip housing and gold wires are bonded from the housing to the contact pads. The layout of the contacts is shown schematically in Fig. 1c and the wire bonded device is shown in a scanning electron micrograph in Fig. 1d, Raman spectroscopy and electrical measurements were performed to verify the presence of graphene (see supporting information).

**ACKNOWLEDGMENTS**

Support from the European Commission through an ERC Advanced Investigator Grant (OSIRIS, No. 228229) and two Starting Grants (M&M's, No. 277879 and InteGraDe, No. 307311) as well as the German Research Foundation (DFG, LE 2440/1-1) and the Italian MIUR through the Cooperlink project (CII11AVUBF) is gratefully acknowledged.

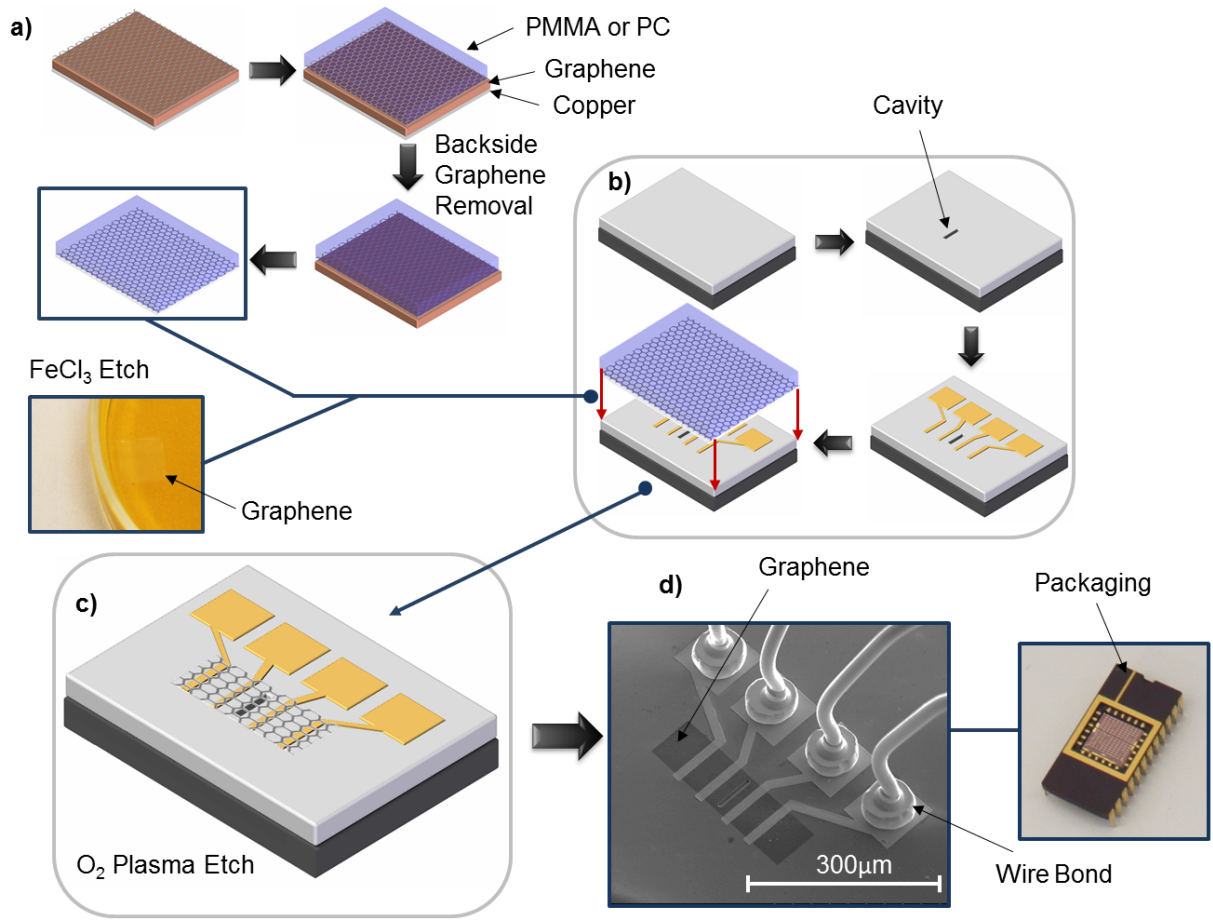

**Figure 1: a)** The graphene transfer process. A Layer of PMMA or PC is applied to one side of chemical vapor deposited (CVD) graphene on copper foil. Graphene is then etched from the back side of the copper foil using $O_2$ plasma. Finally, the copper is etched using $FeCl_3$. **b)** Fabrication sequence of the pressure sensor and the corresponding transfer of graphene onto the substrate. Once the graphene is transferred to the chip, the polymer layer is removed and the graphene is etched (c). After fabrication of the devices, they are packaged and wire bonded (d).

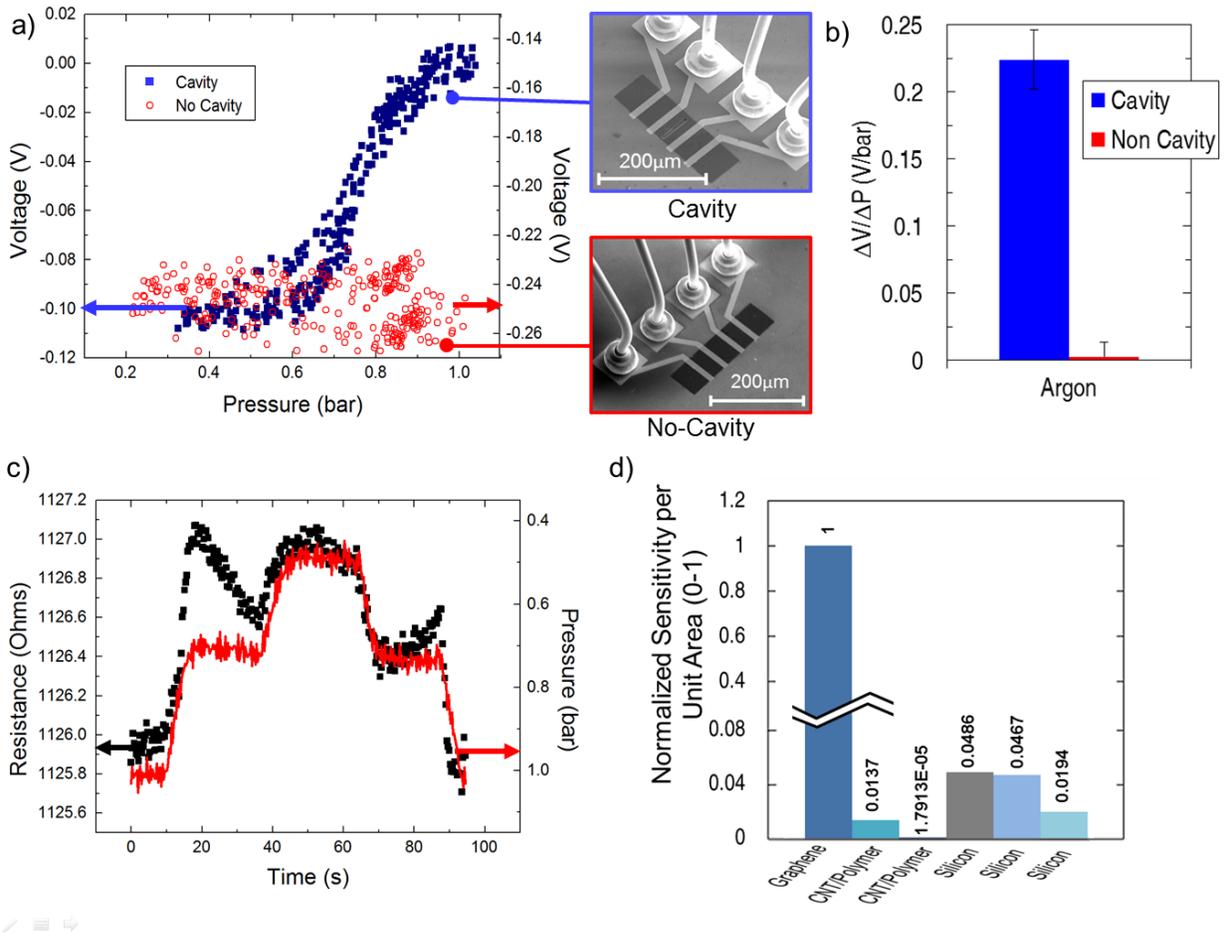

**Figure 2:.** a) Pressure versus voltage measurements of a device with a cavity (blue squares) and a device without a cavity (red hollow circles). There is a clear dependence in the case of the device with a cavity, where the pressure difference leads to bending and strain in the graphene membrane. This dependence is not observed in the unsuspended device. b) Average rate of change of the voltage relative to the pressure for the cavity devices compared to the non-cavity devices. Error bars show their respective standard deviation. c) Resistance of the same cavity device (black squares) compared to the pressure (red line). The pressure was held constant at different levels. d) Comparison of sensitivity. Normalized sensitivity per unit area for the graphene pressure sensors in this **paper compared to silicon and carbon nanotube-based sensors.**[13-16] **The graphene sensor is roughly 20 to 100 times more sensitive per unit area than the conventional MEMS sensors showing the potential for aggressive scaling. Tabulated sensitivity values are shown in the supporting information.**[13-16]

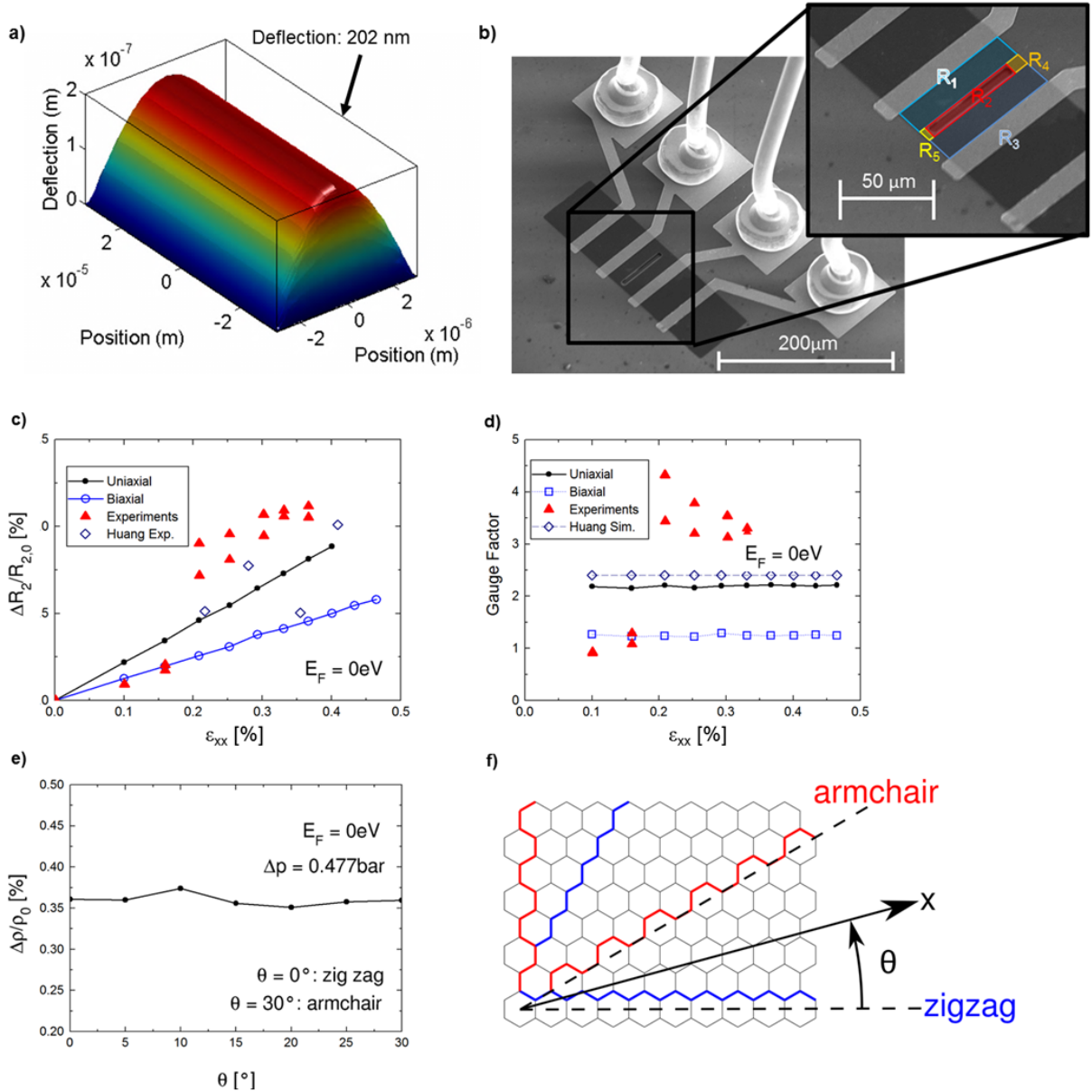

**Figure 3: a)** Model of the cavity deflection for the membrane dimensions in the experiment. The plot shows one half of the symmetric 6 µm by 64 µm membrane in the deflected state at a pressure difference of 477 mbar giving a total deflection of 202 nm. **b)** Different components of the resistor model that was used in order to calculate a gauge factor based on the experimental results. Simulated (lines) and experimental (triangles) relative variation of $R_2$ (c) and the relative gauge

factor (6d) versus strain. Simulation results by Huang et al. added for reference.[19] e) Schematic of the definition of the crystallographic angles used for the calculations in this work.  f) Simulation showing that the strain effect on resistivity is independent of the strain angle.

**Supporting Information Available**

The supporting information contains Raman and electrical data to demonstrate the presence of graphene in the devices as well as details of the Wheatstone bridge setup. It further contains experiments in various gaseous environments to explain how their influence was eliminated and additional measurements of the piezoresistive effect. The material includes a table with the calibration of the COMSOL model with literature data, a detailed explanation of the resistor model we applied and a table comparing the sensitivity of our devices with literature data. Finally, it contains additional figures with results from the tight-binding model. This material is available free of charge via the Internet at http://pubs.acs.org.